\documentclass[aps,prd,twocolumn,nofootinbib,floatfix,superscriptaddress,longbibliography=false]{revtex4-2}

\usepackage[utf8]{inputenc}
\usepackage{amsmath,amssymb}
\usepackage{graphicx}
\usepackage{hyperref}
\usepackage{physics}
\usepackage{booktabs}
\usepackage{xcolor}
\usepackage{svg}
\DeclareMathAlphabet\mathbfcal{OMS}{cmsy}{b}{n}

\begin{document}

\title{CMB Bounds on Primordial Blackholes via Radiation Capture}

\author{M. Farhang}
\email{m\_farhang@sbu.ac.ir}
\affiliation{Department of Physics, Shahid Beheshti University, Tehran 1983969411, Iran}

\author{S. M. S. Movahed}
\affiliation{Department of Physics, Shahid Beheshti University, Tehran 1983969411, Iran}
\affiliation{School of Astronomy, Institute for Research in Fundamental Sciences (IPM), P. O. Box, Tehran 19395-5531, Iran}
\affiliation{Department of Mathematics and Statistics, The University of Lahore, 1-KM Defence Road, Lahore 54000, Pakistan}
\email{m.s.movahed@ipm.ir}

\date{\today}

\begin{abstract}
We explore the capture of neutrinos and photons in the cosmic neutrino and photon background by primordial blackholes (PBHs). 
We model this phenomenon as a gravitational interaction that effectively modifies the continuity equations for radiation and PBH densities and the cosmic expansion history. 
We find that the observability of this modified cosmic history and background densities is highly sensitive to PBH mass,
and only extraordinarily massive PBHs would leave observable trace on the temperature and E-mode polarization of the cosmic microwave background (CMB). 
Specifically,  {\it Planck} data restrict PBH abundance to $f_{\rm pbh}\lesssim 10^{-1}$ for PBH masses above $10^{15} M_\odot$, getting considerably tighter for higher masses. 
We expect substantial improvement as high-resolution measurements of larger CMB multipoles become available. A future cosmic-variance-limited  experiment, with $\ell_{\rm max}=7000$, would set $f_{\rm pbh}\lesssim 10^{-1}-8\times 10^{-5}$ (for the fiducial $\Lambda$CDM cosmology) across $10^{13}-10^{18}M_\odot$. These constraints would  be comparable to the current  limits  at the high-mass end of the spectrum \cite{Carr:2026hot}. 
The gravitational interaction of PBHS with the cosmic background radiation and its imprints on CMB would thus provide an independent complementary probe of extraordinarily massive PBH abundance.
\end{abstract}

\maketitle

\section{Introduction}\label{sec:intro}
Primordial blackholes, or PBHs, are among the most compelling cold dark matter candidates with a rich phenomenology that provides discernible observable predictions. Various PBH imprints, with differing levels of certainty and robustness, have been explored and used to constrain PBH contribution to dark matter, commonly parametrized 
by the current PBH abundance, defined as $f_{\rm pbh}\equiv \rho_{\rm pbh}/\rho_{\rm dm}$. The variety of astronomical and cosmological probes include different physical processes, such as gravitational lensing, dynamical effects, gravitational wave observations, and the impact of gas accretion onto PBHs on the thermal and ionization history of the Universe. 
See, e.g.,  \cite{Josan:2009qn,Carr2010,Carr2021Review,Bellomo:2017zsr,Carr:2017jsz,Sasaki2018,Carr:2020gox,Carr:2020xqk,Green:2020jor} for a review of some of the upper bounds on 
$f_{\rm pbh}$ for various mass scales and mass functions. 
In particular, accretion of baryonic matter onto PBHs at late times can lead to significant energy injection into the surrounding medium, thereby modifying the recombination history and leaving observable imprints on the cosmic microwave background (CMB).

In this work, we explore a physically distinct mechanism that operates predominantly at much earlier times, through gravitational capture of radiation by PBHs. 
Consider a population of uniformly distributed PBHs in a radiation bath deep in the radiation-dominated era, consisting of photons and neutrinos. The photons are tightly coupled to baryons and form a photon-baryon fluid. Neutrinos, on the other hand, are free streaming at nearly the speed of light. 
Neutrinos and photons passing within the corresponding critical impact parameter of a PBH are irreversibly captured and increase the blackhole mass\footnote{It should be noted that certain other processes such as evaporation, baryon accretion and mergers would also lead to PBH mass evolution. See, e.g., \cite{Rice:2017avg, DeLuca:2020bjf,Mosbech:2022lfg,Nesseris:2019fwr,Masina:2020xhk,Custodio:2002gj}. Here we only focus on radiation capture.}.
This would effectively transfer energy from the radiation budget of the cosmos to its matter sector. 
 We model this process by modifying the continuity equations of radiation and dark matter through introducing an effective PBH-radiation interaction, and investigate its impact on the time evolution of dark matter and radiation, and therefore, on cosmic expansion and evolution of cosmological perturbations\footnote{For a complete treatment of the problem, the capture of massive  particles by PBHs should also be taken into account. However, the capture rates are proportional to particle densities, and therefore are most significant at high redshifts. Therefore, we only focus on  relativistic species as the dominant component of the Universe at early times.}.
 We study whether the capture of radiation by PBHs, in the limit where it can be purely modelled  as a geometrical and gravitational phenomenon, would be visible through its potential imprints on CMB anisotropies.
This work can provide a physical motivation for deviations in the evolution of  relativistic species from  standard model predictions\footnote{For a model-independent approach to observational constraints on deviations in the evolution of relativistic species from the predictions of standard model see \cite{Safi:2024bta}.}.

 The interaction mechanism explored in this work is not in principle limited to any 
particular PBH mass range. However, we find that the trace is only significant for extraordinarily massive blackholes, with masses above $10^{13}M_\odot$. This covers a speculative corner of parameter space, with masses far exceeding those typically considered in the standard dark matter window of asteroid- to solar-mass.
Our method thus offers a complementary probe that becomes relevant at the high mass end of the spectrum.
This provides an independent avenue to constrain the abundance of such objects, based on fundamentally different physics than the dynamical, lensing, or accretion-based probes commonly employed in the literature.

\begin{figure}[t]
	\centering 
	\includegraphics[width=0.5\textwidth]{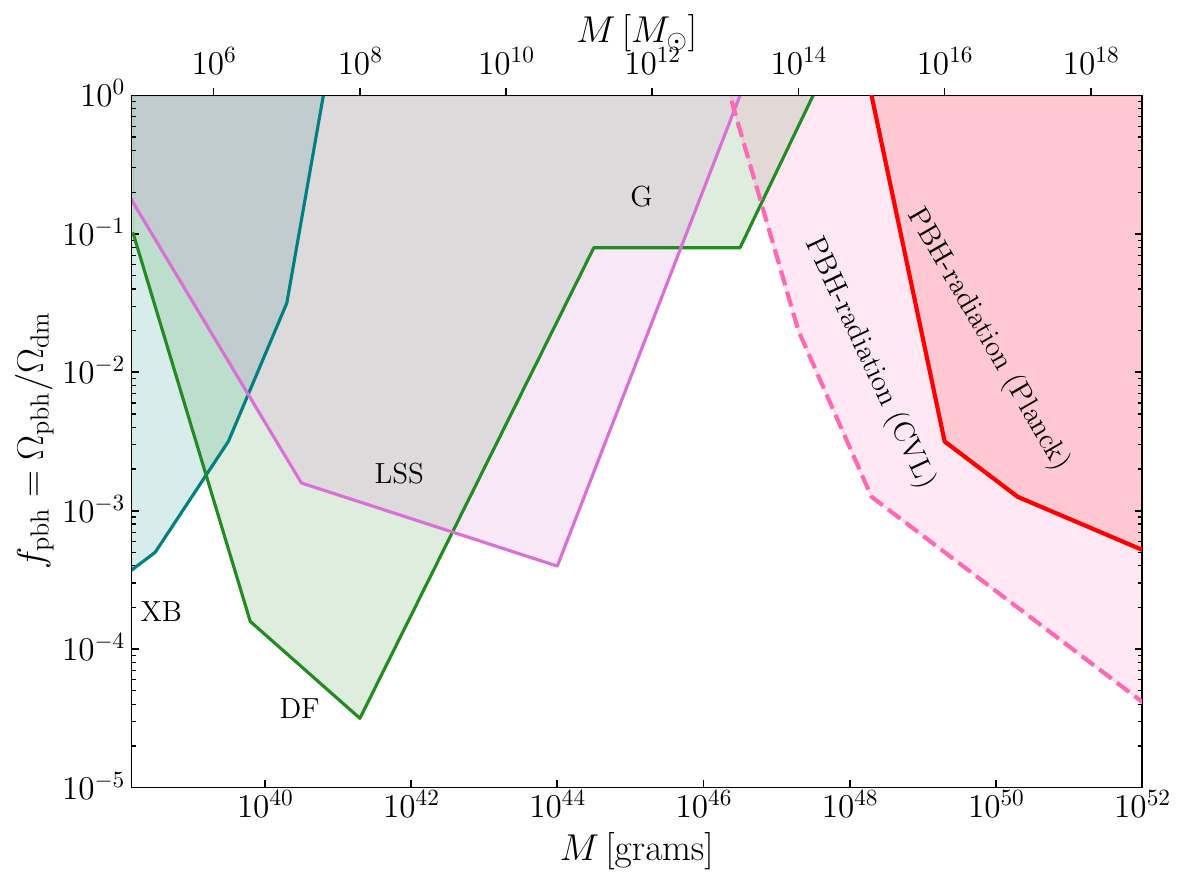}
	\caption{Constraint on PBH abundance based on the PBH-radiation interaction from {\it Planck} temperature and polarization data (combined with BAO), and for a cosmic-variance-limited (CVL) CMB survey with $\ell_{\rm max}=7000$.  For comparison constraints on $f_{\rm pbh}$ for PBHs with very large masses from several other surveys are overplotted (adopted from \cite{Carr:2020erq}), partially based on \cite{Carr:2020xqk}.}
	\label{fig:bounds}%
\end{figure}

The paper is organized as follows. In Section~\ref{sec:nu-pbh} we explain the scenario for PBH-radiation interaction, and in Section~\ref{sec:analysis} discuss the analysis tools to search for their observable imprints. In Section~\ref{sec:results} the implications for PBH abundance are presented and discussed. Figure~\ref{fig:bounds} presents the main results of this paper. We conclude in Section~\ref{sec:conclusion} . Throughout the paper, we assume a monochromatic distribution for PBH mass.
If needed, the fiducial cosmology, unless explicitly stated otherwise, is  assumed to correspond to $\Lambda$CDM scenario with parameters as measured by {\it Planck}  observations \cite{2021A&A...652C...4P}. We work in units with $c=1$.

\section{PBH-radiation interaction}\label{sec:nu-pbh}
We assume PBHs are immersed in a radiation bath, consisting of photons and neutrinos (assumed practically massless at early times of interest here). 
Massless  free-streaming neutrinos follow null geodesics\footnote{Neutrinos decouple from the rest of cosmic plasma  at $z\approx 10^{10}$. For PBHs forming earlier than this epoch, the assumption of free-streaming for neutrinos needs to be modified. However, the PBHs which will be of interest in this work, will form well after neutrino decoupling.}.  
Geometrically, when these geodesics pass sufficiently close to a PBH, the neutrinos would be captured. 
For a Schwarzschild blackhole  with mass $M$
 the critical relativistic impact parameter is $b_{\rm crit}=3\sqrt{3} GM$, and  the capture cross section would thus be
$\sigma_{\nu}=27 \pi (GM)^2$ \cite{Bardeen:1973tla,misner1973gravitation, zakharov1992capture}\footnote{A rigorous treatment of the capture process requires calculating capture cross section for blackholes embedded in an expanding background. The result is expected to be different from Schwarzschild cross section, specially immiditaley after their formation, when the Schwarzschild radius is comparable to the Hubble radius. However, at later times, with the capture process being mainly governed by local physics,  the Schwarzschild  cross section would be a reasonable approximation. As the corrections to capture cross section are confined to the earliest stages of PBH evolution, we expect our main conclusions would not significantly be affected by this approximation.}

Photons, on the other hand, are tightly coupled to baryons before decoupling, forming a relativistic photon-baryon fluid with $c_{\rm s}=1/\sqrt{3}$. Using  Michel model for the spherical accretion of relativistic fluids, the capture cross section is found to be $\sigma_{\gamma}=4 \pi \lambda_{\rm GR}  (GM)^2 c_\infty^{-3}$ 
where $c_\infty$ is the dimensionless sound speed at infinity and $\lambda_{\rm GR}\approx 0.7$ for a radiation-dominated fluid with adiabatic index $\Gamma=4/3$ \cite{1972Ap&SS..15..153M,Aguayo-Ortiz:2021jzv}. We therefore get  $\sigma_{\gamma}\approx 45 \pi (GM)^2$.

The rate of neutrino or photon capture by a single PBH would be $R_{\rm x}=\sigma_{\rm x}n_{\rm x}$,  where  ${\rm x}=\nu,\gamma$ and $n_{\rm x}$ is the  number density of the component.
This capture process is intrinsically local. However, at the homogeneous, background level, it can be effectively modelled as an interaction with the rate of energy exchange 
 described by 
\begin{eqnarray}\label{eq:Q1}
Q_{\rm x}(z)& =&\sigma_{\rm x} \epsilon_{\rm x}(z) n_{\rm x}(z)  n_{\rm pbh}(z) \\ \nonumber \,
&=& A_{\rm x} \pi G^2 M(z) \rho_{\rm x}(z) \rho_{\rm pbh}(z)
\end{eqnarray} 
where $\epsilon_{\rm x}$ represents the average energy of a single particle of type ${\rm x}$, and $\rho_{\rm x}$ and  $n_{\rm x}$ are its average energy and number densities. Also, $M(z)$ is the PBH mass at redshift $z$, $n_{\rm pbh}(z)$ and $\rho_{\rm pbh}(z)$ are average PBH number and energy densities, 
 and  $A_\gamma\approx 45$ and $A_\nu=27$.
This fluid  approximation is analogous to phenomenological treatments of interacting dark matter–dark energy models in the literature, where  the total energy density of the dark sector is conserved but the evolution of the individual component depends on the interaction kernel \cite{Amendola1999er,Valiviita2008iv,Bolotin2013jpa,Wang2016lxa}.
We assume the time evolution of $n_{\rm pbh}$ is solely due to dilution by cosmic expansion.\
Therefore, the  PBH mass evolution, for given current mass $M_{\rm pbh}$ and abundance $f_{\rm pbh}$, can  be described in terms of dark matter and PBH densities as
\begin{equation}\label{eq:Mz}
M(z)=\frac{\rho_{\rm pbh}(z)}{n_{\rm pbh}(z)}=\frac{\rho_{\rm pbh}(z)}{f_{\rm pbh}\rho_{\rm dm,0}(1+z)^3}M_{\rm pbh}
\end{equation}  
where $\rho_{\rm dm,0}$ is the current dark matter density and $M_{\rm pbh}$ is the current PBH mass.   It should be noted that although the evolution of PBH mass is driven by radiation capture which is a highly local process, its spatial variation is expected to remain perturbatively small. In this work, we limit the analysis to the homogeneous component of the energy transfer and its impact on PBH mass evolution. 

The PBH-$\nu$ and PBH-$\gamma$ interactions can be effectively modelled as a decrease in the  neutrino and photon energy densities and a corresponding increase in the 
matter budget of the Universe. Therefore, Eq.~\ref{eq:Q1} can be equivalently considered as the transfer rate of energy density  from radiation into dark matter.
The continuity equations for component x and PBHs with this energy exchange taken into account would thus be
\begin{eqnarray}\label{eq:pbh-nu}
&&\dot{\rho}_{\rm pbh}+3H\rho_{\rm pbh}=Q_\nu+ Q_\gamma\\ \nonumber
&&\dot{\rho}_{\rm x}+3H(1+w_{\rm x})\rho_{\rm x}=-Q_{\rm x}. \label{eq:pbh-nu2}
\end{eqnarray}
Here $w_{\nu,\gamma}=1/3$, and we have modelled PBHs as forming cold dark matter fluid with negligible equation of state.
With the derivatives taken with respect to $z$, and rewriting the equations in  terms of density parameters $\Omega_i(z)\equiv \rho_i(z)/\rho_{\rm c}$  ($i=\nu, \gamma$, pbh), we get
\begin{eqnarray}
\Omega'_{\rm pbh}&=&\frac{3}{1+z}\Omega_{\rm pbh}-\frac{Q_\nu+Q_\gamma}{\rho_c H(1+z)}\\  \nonumber
\Omega'_{\rm x} &=& \frac{3(1+w_{\rm x})}{(1+z)}\Omega_{\rm x}+\frac{Q_{\rm x}}{\rho_c H(1+z)}
\end{eqnarray}	
where prime indicates differentiation with respect to conformal time,  $\rho_{\rm c}$ is the current critical density of the Universe and  we have used Eq.~\ref{eq:Mz} to rewrite $M(z)$ in terms of PBH density parameter. We also have
\begin{equation}\label{eq:Qz}
Q_{\rm x}= A_{\rm x} \pi G^2 \frac{\rho_{\rm c}^2}{\Omega_{\rm dm,0}(1+z)^3}\frac{M_{\rm pbh}}{f_{\rm pbh}} \Omega_{\rm x} \Omega^2_{\rm pbh},
\end{equation}
and the expansion history is described by 
\begin{eqnarray}
H(z)&=&H_0 \Big[ \Omega_{\rm b,0}(1+z)^3 + (1-f_{\rm pbh})\Omega_{\rm dm,0}(1+z)^3\nonumber  \\
&+& \Omega_{\rm pbh}(z)+ \Omega_\gamma (z) + \Omega_\nu (z) + \Omega_\Lambda \Big]^{1/2},
\end{eqnarray}
with subscript $0$ labelling the current value of the parameter.

Finally, in terms of comoving density parameters, defined through $\tilde{\Omega}_{\rm pbh} =\Omega_{\rm pbh}/(1+z)^3$ and $\tilde{\Omega}_{\rm x}=\Omega_{\rm x}/(1+z)^{3(1+w_{\rm x})}$, one gets
\begin{eqnarray} \label{eq:pbh-rad3}
\tilde{\Omega}'_{\rm pbh} &=&-(\tilde{Q}_\nu+\tilde{Q}_\gamma) \frac{(1+z)^{2+3w_{\rm x}}}{ H(z)} \nonumber \\ 
\tilde{\Omega}'_{\rm x} &=& \tilde{Q}_{\rm x} \frac{(1+z)^{2}}{ H(z)}
\end{eqnarray}	
where
\begin{equation} \label{eq:Qtilde}
\tilde{Q}_{\rm x}= \frac{3}{8}A_{\rm x}   \frac{GH_0^2}{\Omega_{\rm dm,0}}\frac{M_{\rm pbh}}{f_{\rm pbh}} \tilde{\Omega}_{\rm x} \tilde{\Omega}^2_{\rm pbh}.  
\end{equation}
Note that $Q_{\rm x}=\rho_{\rm c}\tilde{Q}_{\rm x}(1+z)^{3(2+w_{\rm x})}$. 

\begin{figure}[t]
	\centering 
	\includegraphics[width=0.45\textwidth]{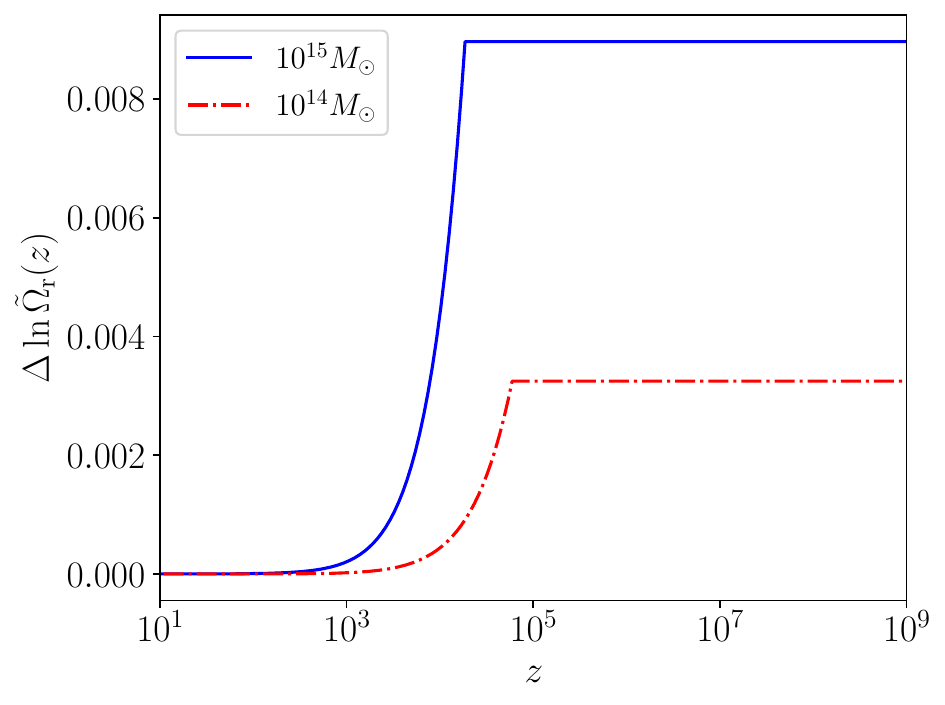}
		\includegraphics[width=0.45\textwidth]{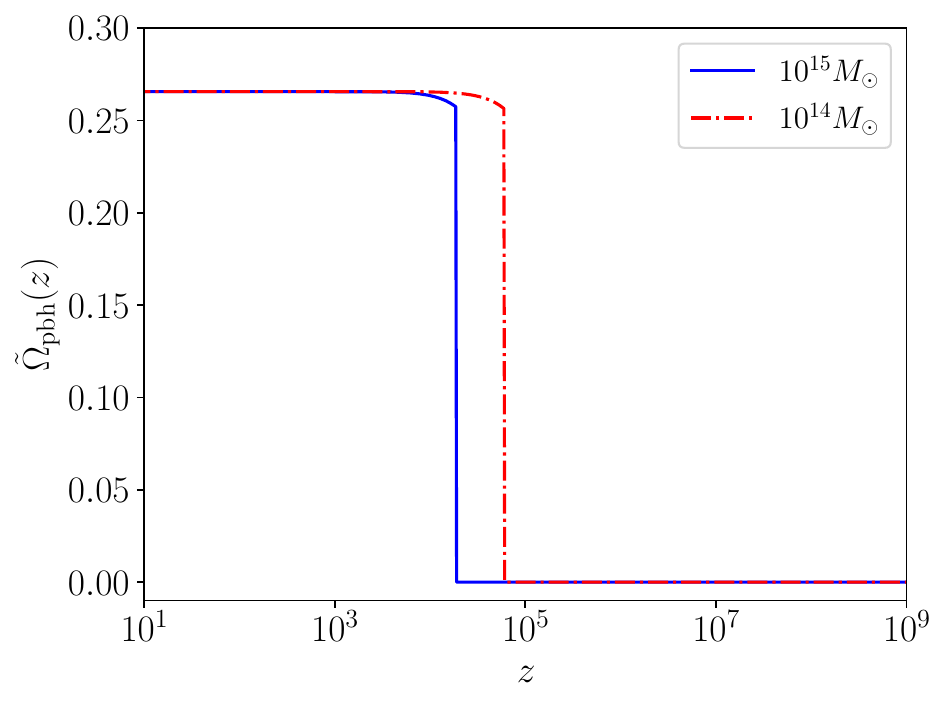}
	\caption{Top: Relative change in comoving radiation density parameter compared to $\Lambda$CDM. Bottom: Comoving PBH density parameter. We have assumed $f_{\rm pbh}=0.1$.}
	\label{fig:omega}%
\end{figure}

\begin{figure}[t]
	\centering 
		\includegraphics[width=0.45\textwidth]{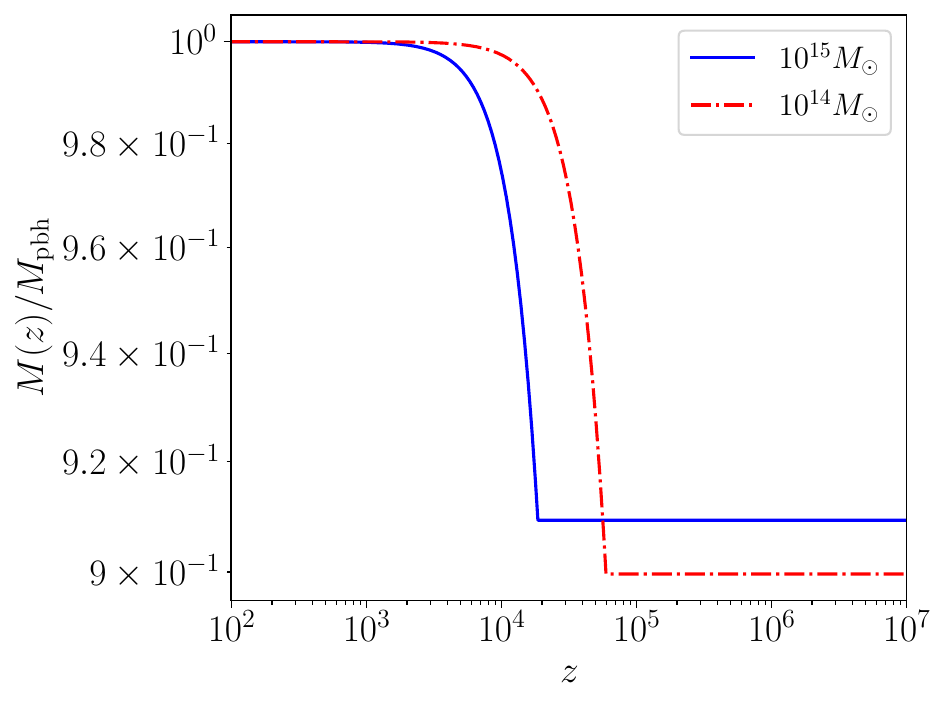}
	\caption{PBH mass evolution due to radiation capture for two PBH masses, normalized to their current masses, as described in Eq.~\ref{eq:Mz}. We have assumed $f_{\rm pbh}=0.1$.}
	\label{fig:M}%
\end{figure}

The interaction would vanish in the limiting case of tiny PBHs (ignoring evaporation for a moment), as the capture cross section depends quadratically on the Schwarzschild radius and thus on  blackhole mass. 
Therefore, as long as the  PBH-radiation interaction matters, with negligible right hand sides for Eqs~\ref{eq:pbh-nu}, \ref{eq:pbh-nu2} and \ref{eq:pbh-rad3}, these PBHs would not be distinguishable from particle dark matter. In this case, $\tilde{\Omega}_{\rm pbh}$ and $\tilde{\Omega}_{\nu,\gamma}$ would remain constant.
It should also be noted that there is no interaction before PBH formation $z_{\rm pbh}$.
Assuming PBHs have formed from the collapse of  large, initial overdensities, $z_{\rm pbh}$ would mark the redshift of horizon entry for 
these fluctuations. It has been shown that the PBH mass would be a faction of the horizon mass at the time of formation, with the proportionality factor scaling in a power-law form with the size of oevrdensity, often found to be $O(0.1)$ \citep{Choptuik:1992jv,Evans:1994pj,Niemeyer:1997mt}.
For our purpose it suffices to take this factor  $\sim 1$.
As a result, $z_{\rm pbh}$ can be approximated for known PBH mass. 
Note that the interaction would then start earlier for smaller masses, as is evident in Figures~\ref{fig:omega} and ~\ref{fig:M} discussed below.
 For simplicity, we model PBH formation as an instantaneous transition in the effective fluid description, transferring the corresponding background energy from the radiation sector to the PBH component. A more realistic treatment would model PBH formation as a smooth transition over a timescale much shorter than the Hubble time, and is therefore not expected to significantly affect the background evolution or cosmological observables.
Also note that the physical interaction rate $Q(z)$ would be 
a highly increasing function of redshift and effectively comes to an end as the radiation and PBH densities dilute with expansion.   


\begin{figure}[t]
	\centering 
	\includegraphics[width=0.5\textwidth]{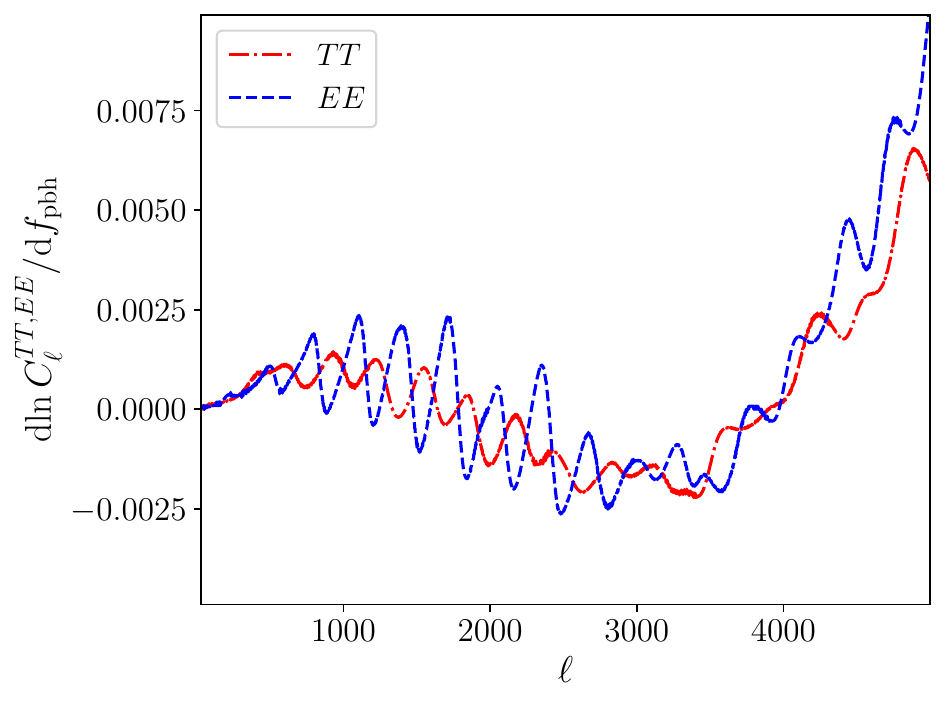}
	\caption{Sensitivity of CMB temperature  and E-mode polarization power spectra to changes in PBH abundance, for $M_{\rm pbh}=10^{14} M_\odot$. The fiducial cosmology is $\Lambda$CDM.}
	\label{fig:Cl}%
\end{figure}

In the $\Lambda$CDM universe, the comoving density parameters are expected to stay constant. Therefore any time evolution of these parameters are due to the energy exchange between PBH and radiation sectors. Figure~\ref{fig:omega} demonstrates the evolution of the comoving radiation and PBH density parameters for $M_{\rm pbh}=10^{14}, 10^{15} M_\odot$ and with $f_{\rm pbh}=0.1$.
The sharp features (i.e., increase in PBH and decrease in radiation comoving densities) mark the epoch of horizon entry of the corresponding fluctuations. 
It is also interesting to see PBH mass evolution due to radiation capture. Figure~\ref{fig:M} illustrates $M(z)$, normalized to unity at $z=0$ for visual aid.

\section{Analysis}\label{sec:analysis}
We modify the Boltzmann solver, CAMB\footnote{\texttt{https://camb.info}} \cite{Lewis:1999bs}, 
by implementing the full time-dependent energy exchange rate, as described in Section~\ref{sec:nu-pbh}, into  background equations which in turn affects the evolution of perturbations.   Besides affecting the expansion history, the change in the photon energy density would alter the  sound speed of the photon-baryon fluid in the tightly coupled regime
and consequently, the sound horizon. These effects have been implemented in the code.

 We note that a more comprehensive description would require treatment of particle capture in an inhomogeneous cosmos by extending the continuity and Euler equations of interacting species to include linear perturbations. The perturbation to the energy transfer term appearing in the continuity equation is relatively straightforward. However, modifying the Euler equations requires a consistent model for the momentum transfer rate which can not be inferred from the background interaction alone. The inclusion of perturbed interactions is expected to give rise to additional observable signatures. We leave this analysis to future work.

We are interested in the imprints of such an interaction on the power spectra of CMB temperature and polarization anisotropies.     
Figure~\ref{fig:Cl} illustrates, for $M_{\rm pbh}=10^{14}M_\odot$, the sensitivity of CMB spectra to PBH abundance (with $\Lambda$CDM as the fiducial cosmology) through PBH-radiation interaction.  The $C_\ell$ response to PBH abundance increases at higher multipoles. 
The decrease in the radiation content of the Universe due to radiation-to-matter transfer would alter expansion history  and consequently, any quantity depending on it. These effects become progressively more pronounced at larger multipoles (corresponding to smaller scales) since they entered the horizon earlier and evolved for longer periods in the  modified  background.
 Therefore, higher resolution CMB surveys are expected to put tighter bounds on $f_{\rm pbh}$. Moreover, the change in the radiation background  alters the  early ISW and radiation driving, which would  affect larger scales as well. The combination of these effects would impact lower multipoles in complex ways.  
 
The final goal is to constrain $f_{\rm pbh}$ with CMB data for various mass scales. 
We add $f_{\rm pbh}$ as a free parameter to the set of six standard cosmological parameters often used to characterize $\Lambda$CDM, i.e., ($\Omega_{\rm b}h^2$, $\Omega_{\rm c}h^2$, $\theta$, $\tau$, $n_{\rm s}$, $A_{\rm s}$), and implement the required changes in the cosmological MCMC code, CosmoMC \cite{Lewis:2002ah}. 
 It should be noted that changes in massless neutrino density is directly  incorporated in density parameter $\Omega_\nu(z)$. This would  consistently impact any background quantity that depends on $\rho_\nu(z)$, and therefore a separate parametrization in the form of $\Delta N_{\rm eff}$ would be unnecessary.

We perform the analysis for different mass scales, as discussed below,  and with {\it Planck} temperature and polarization data \citep{2020A&A...641A..10P}, joint with observations of baryon acoustic oscillations (BAO) \citep{Beutler:2011hx,Ross:2014qpa,BOSS:2016wmc} to help break potential parameter degeneracies.  We also make forecasts for a futuristic CMB survey, with cosmic-variance-limited (CVL) observations up to $\ell_{\rm max}=7000$. 

\section{results and discussion}\label{sec:results}
We use the sensitivity of  CMB power spectra to constrain  PBH abundance through  PBH-radiation interaction as modelled in Section~\ref{sec:nu-pbh}. 
We find this probe to be sensitive to $M_{\rm pbh} \gtrsim 10^{13} M_\odot$, with observable imprints on CMB at lower masses, and the constraints improve considerably as data resolution increases. 
Specifically, the solid red curve in Figure~\ref{fig:bounds} presents  $3\sigma$ upper bounds on $f_{\rm pbh}$, as measured by {\it Planck} (+BAO) data for different $M_{\rm pbh}$'s.
The dashed red curve illustrates the expected constraints with a future CMB experiment assumed to be cosmic-variance limited up to $\ell_{\rm max}=7000$. The main goal is to see how higher multipoles would impact PBH constraints. 
In Figure~\ref{fig:bounds} these CMB-based constraints are compared  against certain astrophysical bounds on {\it stupendously large black holes} as studied by \cite{Carr:2020erq}.
In \cite{Carr:2020erq} the focus has been on dynamical and lensing effects, as well as accretion and gamma ray emission.
However, as is evident from Figure~\ref{fig:bounds}, the mass window explored by the PBH-radiation interaction probe proposed in this work is rather unconstrained by the astrophysical bounds presented in \cite{Carr:2020erq}.  Our MCMC analysis also shows little correlation between $f_{\rm pbh}$ and other standard cosmological parameters, indicating the absence of significant parameter degeneracies. Consequently, the introduction of $f_{\rm PBH}$ has a negligible effect on the inferred constraints of the standard cosmological parameters.

Although blackhole masses with primordial origin and forming in radiation dominated era could be in principle as large as the horizon mass at equality $\sim 10^{17} M_\odot$, PBH masses are not often considered to be this large. 
It should also be noted that in scenarios where PBHs are produced from collapse of $\mathcal{O}(1)$ fluctuations, PBH formation at these mass scales is challenging as CMB observations severely constrain perturbation amplitudes at large scales. 
The constraints derived above are more relaxed than the upper bounds in the high-mass end of PBHs as reported in \cite{Carr:2026hot}. Also
\cite{Carr:2020erq} extensively reviews the astrophysical consequences of {\it stupendously large black holes} (or SLABs), both primordial and those residing in galactic nuclei. The latter would enormously grow in mass due to accretion and one should therefore distinguish between their initial and final masses. 
Nevertheless, it is important to highlight that our proposed method offers a complementary probe by exploring entirely different physical processes, free from astrophysical uncertainty. 
Moreover, the bounds from PBH-radiation interaction are expected to significantly improve with future CVL CMB surveys.
 It should, however, be stressed that a primordial origin for  blackholes in this mass window, should they exist, remains speculative. Nevertheless, several early-Universe mechanisms capable of producing supermassive primordial black holes have been proposed. The resulting mass range, however, is highly model dependent. These mechanisms include the collapse of closed domain walls formed during cosmological phase transitions during inflation \cite{Rubin:2001yw,Khlopov:2004sc}, as well as scenarios based on Affleck-Dine baryogenesis leading to a broad, approximately log-normal PBH mass spectrum extending to supermassive scales \cite{Kawasaki:2019iis}. Inflationary scenarios with enhanced primordial curvature perturbations have also been suggested as possible routes to very massive PBHs or the seeds of supermassive blackholes \cite{Kawasaki:2012kn,Clesse:2015wea}.

It is worth noting that  our treatment applies primarily at early times when relativistic species dominate and the photons and baryons are tightly coupled. This is complementary to, and distinct from late-time baryonic gas accretion \cite{Chen:2016pud,Ali-Haimoud:2016mbv}, particularly effective after recombination,  that injects energy into the plasma and modifies ionization history. Our mechanism of radiation capture, on the other hand, acts as an effective energy sink and removes energy from radiation bath. The two physical processes have therefore different signatures and probe complementary regions of parameter space (of PBH mass) and cosmic history.
This should also be contrasted to the mechanism studied in \cite{Nesseris:2019fwr} where PBH evaporation is modeled as a means of energy transfer from PBH (dark sector) to  radiation.

\section{Conclusion}\label{sec:conclusion}
In this work we explored how neutrino capture and photon accretion by primordial blackholes, if existing, would impact the redshift evolution of radiation and dark matter densities, and therefore the anisotropies of CMB. 
We used this potential imprint to constrain  PBH abundance (Figure~\ref{fig:bounds}).
 While the {\it Planck} data set  $f_{\rm pbh} \lesssim 10^{-1}-5\times 10^{-4}$ for masses in  $(10^{15}-10^{18}) M_\odot$, the expected uncertainties decrease considerably as data quality improves. In particular, we find that with CVL observations up to $\ell_{\rm max}=7000$, the minimum PBH mass with observable PBH-radiation interaction would be as low as $10^{13}M_\odot$, and we get $f_{\rm pbh} \lesssim 10^{-1}-8\times 10^{-5}$ for masses in  $(10^{13}-10^{18}) M_\odot$. 
 
 It is important to stress three  limitations of our results. First, the fluid assumption adopted for PBHs in this work is less evident for PBHs at the high end of the mass spectrum.  
A natural extension of the present framework would be to incorporate the discrete nature of the PBH population, for example through Poisson (shot-noise) fluctuations. We leave a fully self-consistent treatment accounting for these effects for future work.
Moreover, the main focus of the present work is radiation capture by PBHs, which is the dominant component at early times. However, baryonic accretion, which is most relevant for PBHs with the highest masses, introduces additional physical effects, including energy deposition into the surrounding medium, feedback, and modifications to the ionization history, which can also affect CMB observables, including  CMB spectral distortion and anisotropies. 
These processes involve different physical mechanisms from the radiation-capture scenario considered here and have been vastly explored by several groups \citep{Carr:1981pko,Ricotti:2007au, Ricotti:2007jk, Ali-Haimoud:2016mbv,Poulin:2017bwe, Serpico:2020ehh}. In particular, \cite{Ali-Haimoud:2016mbv} find constraints from spectral distortions to be negligible, and constraints  found by \cite{Poulin:2017bwe, Serpico:2020ehh} exclude $f_{\rm pbh}=1$ PBHs with  monochromatic distributions with $M_{\rm pbh} > 2M_\odot $.

 Finally,  the assumption of  steady-state radiation capture requires the characteristic accretion timescale to be much shorter than the Hubble timescale, i.e., $t_{\rm cap}/t_{\rm H} \sim GMH\ll 1$.  
For $M_{\rm pbh} \sim 10^{18} M_\odot$ at around recombination, we get $t_{\rm cap}/t_{\rm H} \sim 0.1-1$ while for $M_{\rm pbh} \sim 10^{15} M_\odot$  the ratio is $\sim  10^{-4}-10^{-3}$. Thus the approximation becomes increasingly more accurate at lower masses, and as PBHs evolve beyond their formation epoch, when the capture process rapidly adjusts to the slowly varying cosmological background. The approximation, however, is least reliable immediately after PBH formation, where a fully time-dependent treatment of the capture process is required.

 The analysis performed in this work is based on the assumption of monochromatic PBH distribution. The work is also limited to the impact of PBH-radiation interaction on the background evolution of cosmic densities and expansion. A  more thorough analysis would require modelling the interaction at the level of (linear) perturbations and the corresponding modification to the Boltzmann equations. We leave these investigations to future work.

\section*{Data Availability}
The results of this work are fully reproducible and the analysis codes are available upon request. 
\section*{ACKNOWLEDGMENTS}
Part of the analysis in this work was performed on the computing cluster of the Canadian Institute for Theoretical Astrophysics.

\bibliography{refs}
\end{document}